\documentclass[aps, prb, twocolumn, showpacs, superscriptaddress]{revtex4-1}

\usepackage{amsmath,amssymb}
\usepackage{graphicx}
\usepackage{dcolumn}
\usepackage{bm}
\begin{document}

\title {A novel topological antiferromagnetic spin-density-wave phase in an extended Kondo lattice model}
\author{Yin Zhong}
\email{zhongy05@hotmail.com}
\affiliation{Center for Interdisciplinary Studies $\&$ Key Laboratory for
Magnetism and Magnetic Materials of the MoE, Lanzhou University, Lanzhou 730000, China}
\author{Yu-Feng Wang}
\affiliation{Center for Interdisciplinary Studies $\&$ Key Laboratory for
Magnetism and Magnetic Materials of the MoE, Lanzhou University, Lanzhou 730000, China}
\author{Yong-Qiang Wang}
\affiliation{Institute of Theoretical Physics, Lanzhou University, Lanzhou 730000, China}
\author{Hong-Gang Luo}
\email{luohg@lzu.edu.cn}
\affiliation{Center for Interdisciplinary Studies $\&$ Key Laboratory for
Magnetism and Magnetic Materials of the MoE, Lanzhou University, Lanzhou 730000, China}
\affiliation{Beijing Computational Science Research Center, Beijing 100084, China}

\date{\today}

\begin{abstract}
By using an extended mean-field theory we study the phase diagram of the topological Kondo lattice model on the honeycomb lattice at half-filling, in which the conduction electrons are described by the Haldane model. Besides the well-defined Kondo insulator and normal antiferromagnetic spin-density-wave (N-SDW) state, it is found that a novel and nontrivial topological antiferromagnetic SDW state (T-SDW) with a quantized Hall conductance is possible if the quasiparticle gap is dominated by the next-nearest neighbor hopping rather than the antiferromagnetic order. By analyzing the low-energy effective Chern-Simon action and the corresponding chiral edge-state, the T-SDW could be considered as a quantum anomalous Hall insulator with antiferromagnetic long-range order. This novel state is apparently beyond Landau-Ginzburg paradigm, which can be attributed to the interplay of quantum anomalous Hall effect and the subtle antiferromagnetic order in the Kondo lattice-like model. While the transition between the SDW states and the Kondo insulator is found to be conventional (a first order transition), the transition between the N- and T-SDWs is, however, a topological quantum phase transition. Interestingly, such topological quantum phase transition can be described by Dirac fermions coupled to a $U(1)$ Chern-Simon gauge-field, which resembles the critical theory between bosonic integer quantum Hall phases and superfluid phase and also indicates such topological quantum phase transition may fall into the $3D$-$XY$ universal class. It is expected that the present work may shed light on the interplay between conduction electrons and the densely localized spins on the honeycomb lattice.
\end{abstract}

\maketitle

\section{Introduction} \label{intr}
Understanding the emergent novel quantum phases and corresponding quantum criticality in heavy fermion compounds is still a challenge in modern condensed matter physics.\cite{Doniach,Sachdev2011,Rosch,Vojta,Custers1,Custers2,Matsumoto,Senthil2003,Senthil2004,Pepin2005,Kim2010,Senthil2010,Zhong2012e}
To capture the essential physics in these phenomena, the Kondo lattice model is introduced to describe the interplay between the Kondo screening and the magnetic interaction, namely, the Ruderman-Kittel-Kasuya-Yosida (RKKY) exchange interaction, which is mediated by conduction electrons among localized spins.\cite{Tsunetsugu} While the former favors a nonmagnetic spin singlet state in strong coupling limit, the latter one tends to stabilize usual magnetic ordered states in weak coupling limit. There seems to exist a quantum phase transition or even a coexistence regime between these two kinds of well-defined states,\cite{Lacroix,Zhang2000,Capponi,Watanabe,Zhang2010,Zhang2011} however, a more radical critical quantum phase perhaps has been observed in the compound of $YbRh_{2}(Si_{0.95}Ge_{0.05})_{2}$, \cite{Custers1,Custers2010} which further motivates people to study more rich phases beyond those mentioned above.

In recent years much progress in fact has been made in exploring the novel quantum phases, \cite{Wen, Saremi, Wen2004, Meng, Clark, Mezzacapo2012, Zhong2012, Hasan2010, Qi2011, Rachel, Hohenadler2011, Ruegg2012, Feng, Mong, Essin, Yoshida, He2011, He2012, Kitaev2008, Schnyder2008, Chen2011a, Chen2011b, Wen2012, Gu2012, Levin2012, Lu2012, Senthil2012,Grover2012,Lu2012b} which is obviously beyond the conventional Landau-Ginzburg paradigm, where the states of matter are classified based on the conventional symmetry-breaking picture. For example, the so-called topological spin-wave-density (T-SDW) state found by He \textit{et al.} \cite{He2011} in the extended Hubbard model can not distinguish from the normal SDW (N-SDW) state according to the broken-symmetry and the corresponding local order parameters introduced, instead He \textit{et al.} used the quantized Hall conductance or the topological matrices $\mathcal{K}$ to identify these novel topological states.\cite{He2011} The success in identifying novel phases by using the quantized Hall conductance motivates us to further explore whether such a novel phase exists or not in other strongly correlated models.

In this work we consider a modified Kondo lattice model, where conduction electrons are described by the Haldane model on the honeycomb lattice at half-filling. Similar to Ref. [\onlinecite{He2011}], this model can be called "topological Kondo lattice". We use the extended mean-field decoupling\cite{Zhang2000} to explore its phase diagram. Besides the well-defined Kondo insulator and the N-SDW phase, a nontrivial T-SDW phase with quantum anomalous Hall effect is possible if the quasiparticle gap is dominated by the next-nearest neighbor hopping rather than the antiferromagnetic order. Furthermore, such novel state can be fully encoded by its low energy effective Chern-Simon action, which underlies the nontrivial quantized Hall response to the external electromagnetic field. Then, by examining the stability of the gapless chiral edge-state derived from the Chern-Simon action, the T-SDW state could be considered as an example of the quantum anomalous Hall insulator with antiferromagnetic long-ranged order.

Moreover, the transition between the mentioned SDW states and the Kondo insulator is found to be conventional first order transition in the sense of the Landau-Ginzburg broken-symmetry picture. However, the transition between the T- and N-SDWs is a topological quantum phase transition. Interestingly, such topological quantum phase transition can be described by two-flavor Dirac fermions coupled to a $U(1)$ Chern-Simon gauge-field, which resembles the critical theory between bosonic integer quantum Hall phases and superfluid phase in Ref. [\onlinecite{Grover2012}] and indicates the critical behaviors should fall into the $3D$-$XY$ universal class. In our view, the similarity between our case and the bosonic integer quantum Hall transition is indeed an interesting new finding and it is desirable to see more examples where some fermionic theories can be dual to certain kinds of bosonic ones.

Additionally, to our knowledge, since no realistic materials could be modeled by the proposed topological Kondo lattice model, we have to expect such model and the novel T-SDW state may be realized in experiments of ultra-cold atoms on the honeycomb optical lattices in near future. The present work has an attempt to uncover novel quantum states beyond the conventional Landau-Ginzburg paradigm for the Kondo lattice-like models on the honeycomb lattice, thus sheds light on the rich physics involved in the heavy fermion systems. The fine but more sophisticated numerical approaches can be used for further study on the system.

The remainder of this paper is organized as follows. In Sec. \ref{sec1}, we first introduce the topological Kondo lattice model on the honeycomb lattice and provide a brief discussion on the Haldane model and its quantized Hall conductivity. Then Sec. \ref{sec2} is devoted to the mean-field treatment of the topological Kondo lattice. Three distinct states are found and one of them is identified as the T-SDW state, which shows the quantum anomalous Hall effect in spite of the established antiferromagnetic long-ranged order. In Sec. \ref{sec3}, the global ground-state phase diagram is proposed based on the mean-field decoupling and the corresponding quantum phase transitions are also discussed. In Sec. \ref{sec4}, the critical theory for the topological quantum phase transition is studied and we find that its critical behaviors should fall into the usual $3D$-$XY$ universal class though the critical theory is formulated by Dirac fermions coupled to $U(1)$ Chern-Simon gauge-field. The Kane-Mele-Kondo lattice model\cite{Feng}, the extended Bernevig-Hughes-Zhang model\cite{Yoshida} and the spin fluctuation effect in SDW states beyond the mean-field treatment of Sec. \ref{sec2} are also briefly discussed in this section. Finally, Sec. \ref{sec5} is devoted to a brief conclusion.

\section{The topological Kondo lattice Model}\label{sec1}
The model we considered is the anisotropic Kondo lattice model, where the conduction electrons are described by the spinful Haldane model on the honeycomb lattice at half-filling,
\begin{eqnarray}
&&H=H_{H}+H_{\parallel}+H_{\perp},\nonumber\\
&&H_{H}=-t\sum_{\langle ij\rangle \sigma}c_{i\sigma}^{\dag}c_{j\sigma}-t'\sum_{\langle\langle ij\rangle\rangle \sigma}e^{i\varphi_{ij}}c_{i\sigma}^{\dag}c_{j\sigma},\nonumber\\
&&H_{\parallel}=\frac{J_{\parallel}}{4}\sum_{i}(c_{i\uparrow}^{\dag}c_{i\uparrow}-c_{i\downarrow}^{\dag}c_{i\downarrow})(d_{i\uparrow}^{\dag}d_{i\uparrow}-d_{i\downarrow}^{\dag}d_{i\downarrow}),\nonumber\\
&&H_{\perp}=\frac{J_{\perp}}{2}\sum_{i}(c_{i\uparrow}^{\dag}c_{i\downarrow}d_{i\downarrow}^{\dag}d_{i\uparrow}+c_{i\downarrow}^{\dag}c_{i\uparrow}d_{i\uparrow}^{\dag}d_{i\downarrow}), \label{eq1}
\end{eqnarray}
where $H_{H}$ is the spinful Haldane model,\cite{Haldane} which supports the quantum anomalous Hall (QAH) effect with two chiral edge states (one for spin-up and the other for spin-down), $t$ and $t'$ are the nearest-neighbor and the next-nearest-neighbor hopping, respectively.\cite{Qi2011} The phase $\varphi_{ij}=\pm\frac{1}{2}\pi$ is introduced to give rise to a quantum Hall effect without external magnetic fields (the so-called QAH) and the positive phase is gained with anticlockwise hopping. Besides, the pseudofermion representation for local spins has been utilized as $S_{i}^{\alpha}=\frac{1}{2}\sum_{\sigma\sigma'}d_{i\sigma}^{\dag}\tau_{\sigma\sigma'}^{\alpha}d_{i\sigma'}$ with $\tau^{\alpha}$ being usual Pauli matrix and a local constraint $d_{i\uparrow}^{\dag}d_{i\uparrow}+d_{i\downarrow}^{\dag}d_{i\downarrow}=1$ enforced in each site. $H_{\parallel}$ denotes the magnetic instability due to the polarization of conduction electrons by local spins while $H_{\perp}$ describes the local Kondo screening effect resulting from spin-flip scattering process of conduction electrons by local moments.

The interplay of the mentioned Kondo screening and the magnetic instability on the honeycomb lattice without the nontrivial next-nearest-neighbor hopping term ($t'$) has been studied by the present authors in the previous work. \cite{Zhong2012b} There either a direct first-order transition or a possible coexistence of the Kondo insulator and the N-SDW state was obtained by the extended mean-field decoupling.

Here, we would like to see whether a novel quantum state could be found in the topological Kondo lattice [Eq. (\ref{eq1})]. This is motivated by the recent work in the so-called topological Hubbard model, where the usual Haldane model is complemented with the Hubbard on-site repulsion interaction $U$.\cite{He2011} In such a model, the T-SDW states with nontrivial edge excitations were discovered and classified by the effective Chern-Simon theory with different $\mathcal{K}$ matrices. Most importantly, all of these T-SDW states have the same physical symmetries with antiferromagnetic long-range order. Thus, in the sense of Landau-Ginzburg paradigm they are the same states. However, these states have different edge states and Hall conductance from the N-SDW states, which display that the T-SDW states are indeed distinguished with their normal counterparts, namely, the N-SDW states.

It should be emphasized that the T-SDW states should not be identified as symmetry-protected-topological (SPT) states\cite{Chen2011a,Chen2011b} since breaking the symmetry of the conservation of total electron number (this symmetry protects the edge-state) will lead to superconducting states rather than a usual antiferromagnetic phase. By original definition in Ref.[\onlinecite{Chen2011a}], the SPT states should smoothly evolve into their corresponding conventional state, e.g. the topological insulator evolves into usual band insulator when breaking the time-reversal symmetry.\cite{Hasan2010,Qi2011} Therefore, based on the above definition, we will not consider the T-SDW states as certain kinds of SPT phases. Meanwhile, it also cannot be considered as the Chern insulator because as a Chern insulator, it should both exhibit a nonzero Hall conductance and preserve the lattice translational symmetry,\cite{Regnault} while the T-SDW states double the effective lattice constant, which obviously breaks original lattice translational symmetry, due to the antiferromagnetic long-range order.

Before moving to the discussion of the topological Kondo lattice model in the next section, it is helpful to give a brief argument on the low energy effective theory of Haldane model and its corresponding Chern-Simon treatment since these issues may not appear in literature of heavy fermions and the same technique will be used in the next section.

\subsection{The massive Dirac fermions from Haldane model}
In this subsection, we will derive an effective action of the Haldane model, which can be described by free massive Dirac fermions in 2+1D.

Our starting point is the spinful Haldane model\cite{Haldane}
\begin{eqnarray}
H_{H}=-t\sum_{\langle ij\rangle \sigma}c_{i\sigma}^{\dag}c_{j\sigma}-t'\sum_{\langle\langle ij\rangle\rangle \sigma}e^{i\varphi_{ij}}c_{i\sigma}^{\dag}c_{j\sigma}. \label{eq2}
\end{eqnarray}
It is useful to rewrite this single-particle Hamiltonian in the momentum space as
\begin{eqnarray}
H_{H}&&=\sum_{k\sigma}-t[f(k)c_{kA\sigma}^{\dag}c_{kB\sigma}+f^{\star}(k)c_{kB\sigma}^{\dag}c_{kA\sigma}] \nonumber\\
&& + 2t'\gamma(k)[c_{kA\sigma}^{\dag}c_{kA\sigma}-c_{kB\sigma}^{\dag}c_{kB\sigma}], \label{eq3}
\end{eqnarray}
where we have defined $f(k)=e^{-ik_{x}}+2e^{ik_{x}/2}\cos(\frac{\sqrt{3}}{2}k_{y})$, $\gamma(k)=\sin(\sqrt{3}k_{y})-2\cos(\frac{3}{2}k_{x})\sin(\frac{\sqrt{3}}{2}k_{y})$ and $A$, $B$ representing two nonequivalent sublattices of the honeycomb lattice, respectively. Then, by diagonalizing the above Hamiltonian, one
obtains the quasiparticle energy band as
\begin{eqnarray}
E_{k\sigma\pm}=\pm\sqrt{t^{2}|f(k)|^{2}+4t'^{2}\gamma(k)^{2}}, \label{eq4}
\end{eqnarray}
which preserves the particle-hole symmetry and also the spin degeneracy. It is well-known that for $3\sqrt{3}t'<t$, the excitation gap mainly opens near six Dirac points (Only two of them are nonequivalent in fact).\cite{Rachel} Then, expanding both $f(k)$ and $\gamma(k)$ near two nonequivalent Dirac points $\pm\vec{K}=\pm(0,\frac{4\pi}{3\sqrt{3}})$, respectively, the gap can be found as $\Delta_{gap}=6\sqrt{3}t'$ and the quasiparticle energy reads $E_{q\sigma\pm}\simeq\pm\sqrt{(\frac{3}{2}tq)^{2}+(3\sqrt{3}t')^{2}}$ with $q=(q_{x},q_{y})\equiv(k_{x},k_{y}\mp\frac{4\pi}{3\sqrt{3}})$.

It is easy to check that both the quasiparticle energy $E_{q\sigma\pm}$ and the excitation gap $\Delta_{gap}$ can be reproduced by the following effective massive Dirac action
\begin{eqnarray}
S_{H}=\int d^{2}xd\tau \mathcal{L}_{0}=\int d^{2}xd\tau\sum_{a\sigma}[\bar{\psi}_{a\sigma}(\gamma_{\mu}\partial_{\mu}+m)\psi_{a\sigma}], \nonumber
\end{eqnarray}
where $\gamma_{\mu}=(\tau_{z},\tau_{x},\tau_{y})$ and $\partial_{\mu}=(\partial_{\tau},\partial_{x},\partial_{y})$
with $\tau_{z},\tau_{x},\tau_{y}$ the usual Pauli matrices. Here the same indices mean summation. We introduce the effective mass $m=-3\sqrt{3}t'$ of Dirac fermions and set the effective Fermi velocity $v_{F}=\frac{3}{2}t$ to unit. The Dirac fields are defined as $\psi_{1\sigma}=(c_{1A\sigma},c_{1B\sigma})^{T}$, $\psi_{2\sigma}=(c_{2A\sigma},-c_{2B\sigma})^{T}$ and $\bar{\psi}_{a\sigma}=\psi^{\dag}_{a\sigma}\gamma_{0}$ with $a=1,2$ denoting the states near the two nonequivalent Dirac points $\pm\vec{K}=\pm(0,\frac{4\pi}{3\sqrt{3}})$ and $T$ implying the transposition manipulation.

\subsection{The Chern-Simon action for the massive Dirac fermions in 2+1D}
Having obtained the effective massive Dirac action in 2+1D, it is interesting to see its physical response to the external electromagnetic field $A_{\mu}=(i\phi,A_{x},A_{y})$ with $\phi$ and $\vec{A}=(A_{x},A_{y})$ representing
the usual scalar and vector potential, respectively. The electromagnetic field can be readily introduced into the Dirac action by the conventional minimal coupling, namely, $\partial_{\mu}\rightarrow\partial_{\mu}-ieA_{\mu}$. Thus, the resulting effective Dirac action coupled with the external electromagnetic field reads
\begin{eqnarray}
S=\int d^{2}xd\tau\sum_{a\sigma}[\bar{\psi}_{a\sigma}(\gamma_{\mu}(\partial_{\mu}-ieA_{\mu})+m)\psi_{a\sigma}].  \label{eq5}
\end{eqnarray}

By integrating out the Dirac fields, we get an effective Chern-Simon action, which represents the electromagnetic response of the massive Dirac fermions to the external electromagnetic field $A_{\mu}$, \cite{He2011}(For details, see Appendix A.)
\begin{eqnarray}
S_{CS}=\int d^{2}xd\tau[Ne^{2}\frac{-im}{8\pi|m|}\epsilon^{\mu\nu\lambda}A_{\mu}\partial_{\nu}A_{\lambda}],  \label{eq6}
\end{eqnarray}
where $N=4$ (two from spins and the other two from the nonequivalent Dirac points), $\epsilon^{\mu\nu\lambda}$ is the usual all-antisymmetric tensor and we have dropped out the regular Maxwell term ($\sim F^{2}_{\mu\nu}$) since the low energy physics is dominated by the Chern-Simon term alone. We should emphasize that although the effective Chern-Simon action is used here, it does not imply any fractionalization or nontrivial topological order (A characteristic signature of the topological order is the ground-state degeneracy depending on the topology of the system.) because no emergent gauge fields or fractionalized quasiparticles exist in the present case.\cite{Wen2004}

Now, one can see that the quantized Hall conductance $\sigma_{H}=\frac{2e^{2}}{h}$ from $J_{x}=\frac{\partial S_{CS}}{\partial A_{x}}|_{\vec{A}\rightarrow0}=Ne^{2}\frac{-im}{4\pi|m|}(\partial_{y}A_{0}-\partial_{0}A_{y})=\frac{2e^{2}}{2\pi}E_{y}$
where $h=2\pi\hbar$ is reintroduced and we have also used $m/|m|=-1$ and $N=4$.\cite{Wen2004} Since the quantized Hall conductance is realized without any external magnetic field ($\vec{A}\rightarrow0$), the Haldane model (or its low energy theory) just provides an example of the quantum anomalous Hall effect, which is defined as a band insulator with quantized Hall conductance but without orbital magnetic field.\cite{Qi2011}

\section{Mean-field treatment of topological Kondo lattice} \label{sec2}
In this section, we will use mean-field decoupling to study the ground-state phase diagram of the topological Kondo lattice model [Eq. (\ref{eq1})]. Fluctuation effect beyond the present mean-field treatment will be analyzed in Sec. \ref{sec4}.

By utilizing the mean-field decoupling introduced by Zhang and Yu \cite{Zhang2000} for the longitudinal and transverse interaction term $H_{\parallel}$, $H_{\perp}$, respectively, it is straightforward to obtain a mean-field Hamiltonian
\begin{eqnarray}
&&H_{MF}=H_{H}+H_{\parallel}^{MF}+H_{\perp}^{MF}+E_{0},\nonumber\\
&&H_{\parallel}^{MF}=\frac{J_{\parallel}}{2}\sum_{k\sigma}[\sigma(-m_{c}d_{kA\sigma}^{\dag}d_{kA\sigma}+m_{d}c_{kA\sigma}^{\dag}c_{kA\sigma})\nonumber\\
&& \hspace{2cm} - (A\rightarrow B)], \nonumber\\
&&H_{\perp}^{MF}=\frac{J_{\perp}V}{2}\sum_{k\sigma}(c_{kA\sigma}^{\dag}d_{kA\sigma}+c_{kB\sigma}^{\dag}d_{kB\sigma}+h.c.),\nonumber\\
&&E_{0}=N_{s}(2J_{\parallel}m_{d}m_{c}+J_{\perp}V^{2}), \label{eq7}
\end{eqnarray}
where $N_s$ is the number of lattice sites. We have defined several mean-field parameters as $\langle d_{iA\uparrow}^{\dag}d_{iA\uparrow}-d_{iA\downarrow}^{\dag}d_{iA\downarrow}\rangle = 2m_{d}$,
$\langle d_{iB\uparrow}^{\dag}d_{iB\uparrow}-d_{iB\downarrow}^{\dag}d_{iB\downarrow}\rangle = -2m_{d}$, $\langle c_{iA\uparrow}^{\dag}c_{iA\uparrow}-c_{iA\downarrow}^{\dag}c_{iA\downarrow}\rangle = -2m_{c}$,
$\langle c_{iB\uparrow}^{\dag}c_{iB\uparrow}-c_{iB\downarrow}^{\dag}c_{iB\downarrow}\rangle = 2m_{c}$ and
$-V=\langle c_{i\uparrow}^{\dag}d_{i\uparrow}+d_{i\downarrow}^{\dag}c_{i\downarrow}\rangle =
\langle c_{i\downarrow}^{\dag}d_{i\downarrow}+d_{i\uparrow}^{\dag}c_{i\uparrow}\rangle$. It can be seen that $m_{d}, m_{c}$ correspond to magnetization of local spins and conduction electrons, respectively, while non-vanishing $V$ denotes
the onset of Kondo screening effect. Besides, since we are considering a half-filled lattice, the local constraint of the pseudofermion has been safely neglected at the present mean-field level with chemical potential setting to zero.\cite{Zhang2000}

Firstly, we proceed to discuss two simple but physically interesting limits for Kondo coupling $J_{\parallel}$ and $J_{\perp}$, which correspond to the antiferromagnetic SDW state ($J_{\parallel}\gg J_{\perp}$) and Kondo insulating state ($J_{\parallel}\ll J_{\perp}$), respectively.\cite{Tsunetsugu}

\subsection{The antiferromagnetic spin-density-wave state}
For the case with $J_{\parallel}\gg J_{\perp}$, in general, one expects that the antiferromagnetic SDW state to be the stable ground-state of Kondo lattice model on the honeycomb lattice due to its bipartite feature.\cite{Tsunetsugu} To study the possible antiferromagnetic ordered state, diagonalizing the mean-field Hamiltonian (\ref{eq7}) with assuming no Kondo screening existing ($V=0$), we can easily derive ground-state energy of the antiferromagnetic SDW state per site as
\begin{eqnarray}
E_{g}^{AFM}&&=J_{\parallel}m_{c}(2m_{d}-1) \nonumber\\
&&-\frac{1}{N_{s}}\sum_{k\sigma}\sqrt{(\frac{\sigma J_{\parallel}m_{d}}{2}+\gamma(k))^{2}+t^{2}|f(k)|^{2}}\nonumber
\end{eqnarray}
and two self-consistent equations from minimizing $E_{g}^{AFM}$ with respect to magnetization $m_{d}$ and $m_{c}$, respectively.
\begin{eqnarray}
&&J_{\parallel}m_{c}(2m_{d}-1)=0,\nonumber\\
&&m_{c}=\frac{1}{4N_{s}}\sum_{k\sigma}\frac{J_{\parallel}m_{d}/2+\sigma\gamma(k)}{\sqrt{(\frac{\sigma J_{\parallel}m_{d}}{2}+\gamma(k))^{2}+t^{2}|f(k)|^{2}}}.\nonumber
\end{eqnarray}

For further analytical treatment, we may use a simplified linear density of state (DOS) $\rho(\varepsilon)=|\varepsilon|/\Lambda^{2}$ when transforming the summation over momentum $k$ into integral on energy $\varepsilon$ with $\Lambda\simeq2.33t$ being high-energy cutoff. Thus, $t|f(k)|$ can be replaced by $|\varepsilon|$ to simplify corresponding calculations and $\gamma(k)$ is replaced by $\pm3\sqrt{3}t'$ near two nonequivalent Dirac points $\pm\vec{K}=\pm(0,\frac{4\pi}{3\sqrt{3}})$.

From these two equations, one obtains $m_{d}=1/2$ and
\begin{equation}
m_{c}=\frac{1}{2\Lambda^{2}}\sum_{\sigma} R_\sigma (\sqrt{\Lambda^{2}+R_\sigma^{2}}-R_\sigma)\nonumber
\end{equation}
with $R_\sigma = \sigma3\sqrt{3}t'+J_{\parallel}/4$
while the ground-state energy per site for the antiferromagnetic SDW state reads
\begin{equation}
E_{g}^{AFM}=-\frac{2}{3\Lambda^{2}}\sum_{\sigma}[(\Lambda^{2}+R_\sigma^{2})^{3/2}-R_\sigma^3].\label{eq8}
\end{equation}

\begin{figure}
\includegraphics[width=0.9\columnwidth]{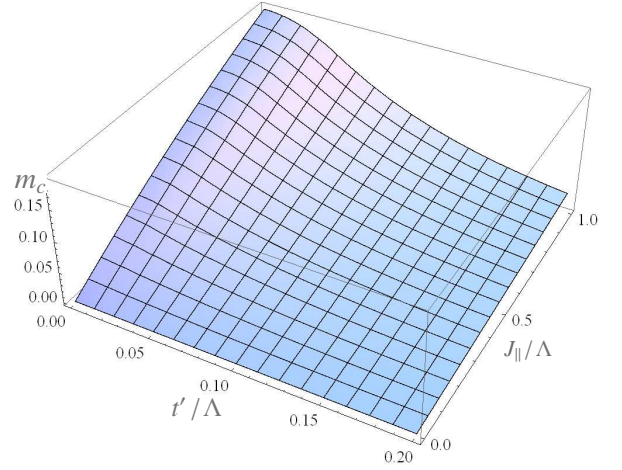}
\caption{\label{fig:1} Magnetization of conduction electrons ($m_{c}$) versus next-nearest-neighbor hopping $t'$ and the Kondo coupling $J_{\parallel}$ in the antiferromagnetic SDW state with $\Lambda\simeq2.33t$ being high-energy cutoff.}
\end{figure}

Apparently, the local spins are fully polarized ($m_{d}=1/2$) while the conduction electrons have small magnetization as shown in Fig.~\ref{fig:1}. It is also noted that when $t'=0$ (no next-nearest-neighbor hopping), the above $m_{c}$ correctly recovers the value in our previous work.\cite{Zhong2012b} Meanwhile, the low-lying quasiparticle excitations in the antiferromagnetic SDW state has the energy $E_{\pm\sigma}^{1,2}(k)=\pm\sqrt{(3tk/2)^{2}+(\sigma J_{\parallel}/4\pm3\sqrt{3}t')^{2}}$ and $E_{\pm\sigma}^{3,4}(k)=\pm J_{\parallel}m_{c}/2$. It should be noted that the gap around the Dirac points only closes when the condition $J_{\parallel}/4=3\sqrt{3}t'$ is fully satisfied, otherwise, any low-lying quasiparticle excitations in the antiferromagnetic SDW state are clearly gapped. Thus, we may conclude that the antiferromagnetic SDW state we obtained is mainly an insulating state (except for the case with $J_{\parallel}/4=3\sqrt{3}t'$) with fully polarized local spins ($m_{d}=1/2$) while conduction electrons only partially polarize ($m_{c}<1/2$). This feature is similar to the previous study on square lattice, thus confirms the validity of our current treatment.\cite{Zhang2000}

Additionally, as a matter of fact, with the help of a low energy effective theory similar to the one in Sec. \ref{sec1}, the mentioned case for vanished gap ($J_{\parallel}/4=3\sqrt{3}t'$) can be identified as a topological quantum phase transition between a N-SDW state and a T-SDW one, which shows a quantum anomalous Hall effect in spite of the antiferromagnetic long-ranged order. More details will be pursued in the next subsection.

\subsection{Topological quantum phase transition and the T-SDW state}
After obtained the condition for the vanishing gap in the last subsection, it is interesting to see what new physics this will lead to and whether a novel SDW state may be uncovered.

Performing the same treatment as for the Haldane model in Sec.\ref{sec1} on the mean-field Hamiltonian  (\ref{eq7}),  we obtains the following effective action for the antiferromagnetic SDW state
\begin{eqnarray}
S=\int d^{2}xd\tau\sum_{a\sigma}[\bar{\psi}_{a\sigma}(\gamma_{\mu}(\partial_{\mu}-ieA_{\mu})+m_{a\sigma})\psi_{a\sigma}],\label{eq9}
\end{eqnarray}
where the effective mass is defined as $m_{1\uparrow}=m_{2\downarrow}=m-J_{\parallel}/4$ and $m_{1\downarrow}=m_{2\uparrow}=m+J_{\parallel}/4$ with $m=-3\sqrt{3}t'$. Then, it is straightforward to derive an effective Chern-Simon action by integrating out the Dirac fermions
\begin{eqnarray}
S_{CS}=\int d^{2}xd\tau[We^{2}\frac{-i}{8\pi}\epsilon^{\mu\nu\lambda}A_{\mu}\partial_{\nu}A_{\lambda}]\nonumber
\end{eqnarray}
and we have also defined $W=\sum_{a\sigma}\frac{m_{a\sigma}}{|m_{a\sigma}|}$. It is easy to see that a quantized Hall conductance with the value $\sigma_{H}=2e^{2}/h$ ($W=-4$) is obtained if $3\sqrt{3}t'>J_{\parallel}/4$ (See the second subsection of Sec.\ref{sec1} for the calculation of the quantized Hall conductance.). In contrast, when $3\sqrt{3}t'$ is smaller than $J_{\parallel}/4$, no such quantized Hall conductance can be found ($W=0$) and the corresponding effective Chern-Simon term vanishes.

\begin{figure}
\includegraphics[width=0.9\columnwidth]{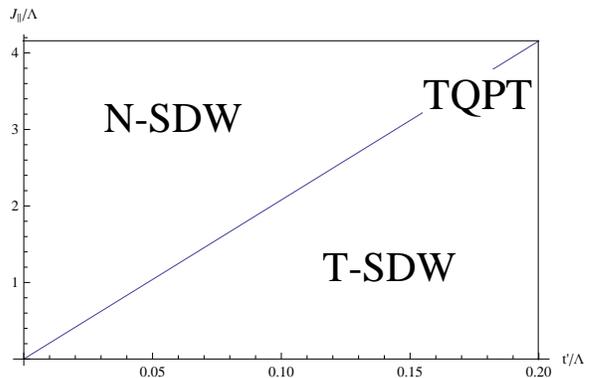}
\caption{\label{fig:2} The topological quantum phase transition (TQPT) between the normal antiferromagnetic SDW (N-SDW) state and the topological antiferromagnetic SDW (T-SDW) state with quantum anomalous Hall effect. The boundary of these two kinds of states is determined by $3\sqrt{3}t'=J_{\parallel}/4$ with $\Lambda\simeq2.33t$ being high-energy cutoff.}
\end{figure}

Therefore, it seems that even in the antiferromagnetic SDW state, there exists a quantized Hall conductance without external magnetic fields, if the quasiparticle gap is still dominated by the next-nearest neighbor hopping ($3\sqrt{3}t'$) rather than the antiferromagnetic order ($J_{\parallel}/4$). Thus, we have uncovered a topological antiferromagnetic SDW state, namely, the T-SDW state with a quantum anomalous Hall effect for $3\sqrt{3}t'>J_{\parallel}/4$ (Recalling that the quantum anomalous Hall effect is defined as a band insulator with quantized Hall conductance but without orbital/external magnetic field.).

It is also interesting to see that in the T-SDW state, one may use the following effective Chern-Simon action to reproduce the quantized Hall conductance
\begin{equation}
S=\int d^{2}xd\tau[K_{IJ}\frac{-i}{4\pi}\epsilon^{\mu\nu\lambda}a_{I\mu}\partial_{\nu}a_{J\lambda}+\frac{ie}{2\pi}q_{I}\epsilon^{\mu\nu\lambda}A_{\mu}\partial_{\nu}a_{I\lambda}],\nonumber
\end{equation}
where we have defined the so-called $K$ matrix as $K_{IJ}=\delta_{IJ}$ for $I,J=1,2$ and the corresponding charge vector reads $q=(1,1)^{T}$. Physically, the effective gauge field $a_{1\mu}$ and $a_{2\mu}$ are introduced to give rise to the conserved current for spin-up and spin-down electrons, respectively. (So the system has $U(1)\times U((1)$ symmetry for two kinds of spins and a global $U(1)$ symmetry for the conservation of total electron number (charge) as well.) For the physical observable, the filling factor $\nu$, which determines the quantized Hall conductance as $\sigma_{H}=\nu\frac{e^{2}}{h}$, can be calculated by $\nu=q^{T}K^{-1}q=2$, thus the effective action correctly reproduces the quantized Hall conductance obtained in the previous paragraph. Moreover, since $Det[K]=1$, no states with topological order and fractional excitations are involved\cite{Lu2012} (Generally, fractionalized excitations require a $K$-matrix with $|Det[K]|>1$ since the ground-state degeneracy on a torus is equal to $Det[K]$.\cite{Senthil2012}). In fact, the elementary quasiparticle
(the electron) can also be obtained in the present formulism by checking the so-call exchange statistical angle $\theta$. Without any surprise, we find such angle is $\pi$ which means that the exchange of two identical quasiparticle give rise to a $\pi$ phase in their wavefunction. Therefore, the elementary quasiparticle is the usual electron as expected.

To deepen the discussion, one may utilize the bulk-edge correspondence for the above effective abelian Chern-Simon theory to derive two decoupled gapless chiral edge states as follows
\begin{eqnarray}
S_{edge}=\sum_{I=1,2}\int dxd\tau\frac{1}{4\pi}[-i\partial_{\tau}\phi_{I}\partial_{x}\phi_{I}+c_{I}\partial_{x}\phi_{I}\partial_{x}\phi_{I}]\nonumber
\end{eqnarray}
with $c_{I}$ denoting the non-universal velocity of edge states and $\phi_{I}$ being the bosonic representation for the two edge-state modes ($I=1$ for the spin-up mode while $I=2$ for the spin-down mode). Because each edge-state mode contributes $e^{2}/h$ to the Hall conductance, the total quantized Hall conductance is obviously $2e^{2}/h$ as expected. It is noted that the above quantized value for Hall conductance is protected by the chiral feature of the edge-state even when the conservation of electron number with different spins is broken (Such breaking could happen
when fluctuations beyond the present mean-field-like treatment are considered and the stability of the gapless chiral edge-stat is discussed in Appendix B.). In contrast, if one breaks the $U(1)$ symmetry for conservation of total electron number (e.g. by some superconducting paring terms), the T-SDW state cannot be stable anymore since its quantized Hall conductance (or the gapless chiral edge-state) will be destroyed generically. (Our case is similar to the bosonic example provided by Senthil and Levin very recently,\cite{Senthil2012} where the conservation of total boson number protects the quantized (charge) Hall conductance.) However, the T-SDW state may not be considered as a kind of the symmetry-protected-topological states\cite{Chen2011a,Chen2011b} since breaking the conservation of total electron number only leads to some superconducting states but not the expected usual antiferromagnetic SDW states.

However, one finds that there is only a usual antiferromagnetic SDW state, namely, the N-SDW when the gap is dominated by the antiferromagnetic order ($J_{\parallel}/4>3\sqrt{3}t'$). Obviously, these two SDW states have the same physical symmetry and at the same time break the spin-rotation invariance. However, they are rather different states of matter  due to the response to the electromagnetic field (The former has quantized Hall conductance while the latter one does not.). Therefore, this provides a simple example beyond the Landau-Ginzburg paradigm, particularly the symmetry-breaking based classification of states of matter.\cite{Wen2004} We note that although the T-SDW state and the N-SDW states cannot be fully identified by symmetry-breaking paradigm, one may use the their distinct quantized Hall conductance (or their low energy effective Chern-Simon action and the robust chiral edge states) to identify and classify them completely.

Moreover, when the gap closes, a topological quantum phase transition similar to the one in Ref. [\onlinecite{He2011}] appears between these two distinct antiferromagnetic SDW states and gapless Dirac fermions reappear near the Dirac points. However, in contrast to the case in Ref. [\onlinecite{He2011}], no discontinuity of the third derivative of the ground-state energy is found near the topological quantum phase transition. This is due to the fact that in our case the effective magnetic field, which the conduction electrons experience, is fixed ($m_{d}=1/2$) throughout the whole antiferromagnetic region while the effective magnetic field varies in Ref. [\onlinecite{He2011}]. More importantly, when considering fluctuation effect in Sec. \ref{sec4}, we will find that the mentioned topological quantum phase transition should fall into the $3D$-$XY$ universal class though the critical theory is fermionic. We note that such new feature is not reported in Ref. [\onlinecite{He2011}]. The main findings in this subsection are systematically summarized in Fig.~\ref{fig:2}.

\subsection{The Kondo insulating state}
Another interesting case appears when $J_{\parallel}\ll J_{\perp}$. It is natural to expect that a Kondo insulating  state arises in this situation for half-filling.\cite{Tsunetsugu,Zhang2000} Following the same methology of treating antiferromagnetic SDW state, we can get the ground-state energy per site for the expected Kondo insulating state with $V\neq0$ but no magnetic orders $m_{d}=m_{c}=0$
\begin{eqnarray}
E_{g}^{Kondo}&&=J_{\perp}V^{2}-\frac{4}{3\Lambda^{2}}[(\Lambda^{2}+(3\sqrt{3}t')^{2}+J_{\perp}^{2}V^{2})^{3/2}\nonumber\\
&&-((3\sqrt{3}t')^{2}+J_{\perp}^{2}V^{2})^{3/2}].\label{eq10}
\end{eqnarray}

Minimizing $E_{g}^{Kondo}$ with respect to Kondo hybridization parameter $V$, we obtain $V^{2}=\frac{1}{16J_{\perp}^{4}}[\Lambda^{4}-8\Lambda^{2}J_{\perp}^{2}-16J_{\perp}^{2}(3\sqrt{3}t')^{2}+16J_{\perp}^{4}]$ which implies a critical coupling $J_{\perp}^{c}=\frac{2}{\Lambda^{2}}[\sqrt{\Lambda^{2}+(3\sqrt{3}t')^{2}}-3\sqrt{3}t']$ corresponding to vanishing $V$. It is noted that $V\propto(J_{\perp}-J_{\perp}^{c})$ in contrast to usual mean-field result $\beta=1/2$, and this can be attributed to the low-energy linear DOS of conduction electrons on the honeycomb lattice at half-filling.
Similar critical behavior for onset of Kondo screening on the honeycomb lattice has been obtained in the study of Kondo breakdown mechanism as well.\cite{Saremi} As a matter of fact, the existence of the critical coupling $J_{\perp}^{c}$ results from the competition between the Kondo insulating state and the nontrivial decoupled state [This decoupled state is in fact the quantum anomalous Hall state described by the spinful Haldane model in Eq. (\ref{eq2}).] where $V=m_{d}=m_{c}=0$, its ground-state energy $E_{g}^{0}=-\frac{4}{3\Lambda^{2}}[(\Lambda^{2}+(3\sqrt{3}t')^{2})^{3/2}-(3\sqrt{3}t')^{3}]$ comes solely from free conduction electrons. Comparing $E_{g}^{0}$ and $E_{g}^{Kondo}$, one clearly recovers the critical coupling $J_{\perp}^{c}$, which justifies the above simple picture.

However, since the Kondo insulating state is unstable to the decoupled state when $J_{\perp}<J_{\perp}^{c}$, one may wonder whether a nontrivial decoupled state appears between the Kondo insulating state and the antiferromagnetic SDW states. It is easy to see that the ground-state energy of the antiferromagnetic SDW state $E_{g}^{AFM}$ is always lower than $E_{g}^{0}$ for any positive coupling $J_{\parallel}$ and the next-nearest-neighbor hopping $t'$. Therefore, a nontrivial quantum anomalous Hall state in intermediate coupling seems unfavorable based on our current mean-field treatment. However, one notes that the T-SDW state found in the previous subsection in the antiferromagnetic ordered region shows the quantum anomalous Hall effect.

\section{The global phase diagram of topological Kondo lattice} \label{sec3}
Having found the antiferromagnetic SDW states (both the N- and T-SDW states) and the kondo insulator in the previous section by mean-field decoupling, here we proceed to discuss the possible ground-state phase diagram.

\begin{figure}
\includegraphics[width=0.9\columnwidth]{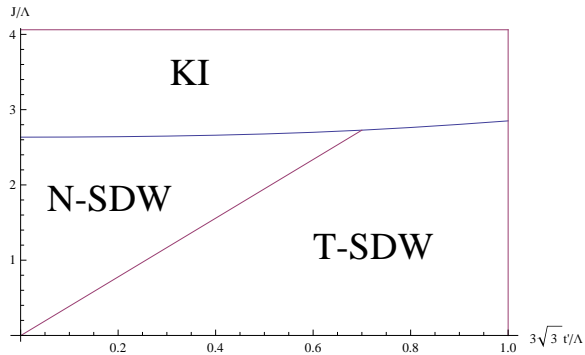}
\caption{\label{fig:3} The ground-state phase diagram of the topological Kondo lattice model. KI is the Kondo insulator and the quantum phase transition between the KI and the antiferromagnetic SDW states is first order. In the antiferromagnetic SDW region, N-SDW denotes the normal antiferromagnetic SDW state while the T-SDW represents the topological antiferromagnetic SDW state with quantum anomalous Hall effect. The boundary of these two kinds of SDW states is determined by $3\sqrt{3}t'=J_{\parallel}/4$ with $\Lambda\simeq2.33t$ being high-energy cutoff.}
\end{figure}

Since a nontrivial decoupling state (the quantum amomalous Hall state) is excluded in the last subsection,
instead, we may consider possible second-order quantum phase transition between Kondo insulating state and the antiferromagnetic SDW states. The boundary of the transitions can be determined by comparing the ground-state energies of $E_{g}^{AFM}$ [Eq. (\ref{eq8})] and $E_{g}^{Kondo}$ [Eq. (\ref{eq10})]. For physically interesting case with $J_{\perp}=J_{\parallel}=J$, the phase diagram is shown in Fig.~\ref{fig:3}, from which we can see that there exist three distinct states, one is the Kondo insulator (KI) and the other two are the N-SDW state and the T-SDW state with quantum anomalous Hall effect, which is stable for large next-nearest-neighbor hopping $t'$. The phase transition between the two kinds of SDW states is the topological quantum phase transition which has been studied in previous section. However, in contrast to one's expectation, one can check that these putative continuous second-order quantum phase transition between the Kondo insulating state and the antiferromagnetic SDW states are in fact first-order when comparing the first-order derivative of $E_{g}^{Kondo}$ and $E_{g}^{AFM}$ with respect to the Kondo coupling $J$ for different fixed $t'$. Thus, we do not expect radical critical behaviors near such first-order quantum phase transition points in the spirit of Landau-Ginzburg paradigm.\cite{Sachdev2011}

Additionally, it is noted that the mentioned first-order quantum phase transition has also been obtained on the square lattice\cite{note} (with $t'=0$) and we suspect this feature may be generic for conventional mean-field treatment of Kondo lattice models according to standard Landau-Ginzburg phase transition theory.\cite{Zhang2000} However, we point out that it is a subtle issue to compare the results of a first-order quantum transition with numerical simulations, particularly when the first-order transition is a weak one.\cite{Continentino} Moreover, generically, a possible coexistence region of the Kondo insulating state and the antiferromagnetic ordered state cannot be excluded but we will leave this interesting issue for future work.

\section{Extension and Discussions} \label{sec4}
\subsection{Spin fluctuation effect in antiferromagnetic spin-density wave states}
Here, we discuss the spin fluctuation effect in the antiferromagnetic spin-density wave states, which is omitted in the previous mean-field treatment in Sec. \ref{sec2}.

First, let us recall the effective Dirac action [Eq. (\ref{eq9})]
\begin{eqnarray}
S=\int d^{2}xd\tau\sum_{a\sigma}[\bar{\psi}_{a\sigma}(\gamma_{\mu}(\partial_{\mu}-ieA_{\mu})+m_{a\sigma})\psi_{a\sigma}],\nonumber
\end{eqnarray}
where we have defined the effective mass as $m_{1\uparrow}=m_{2\downarrow}=m-J_{\parallel}/4$ and $m_{1\downarrow}=m_{2\uparrow}=m+J_{\parallel}/4$ with $m=-3\sqrt{3}t'$.

Following Ref. \onlinecite{He2011}, when the antiferromagnetic order is well established, the effect of spin fluctuations is equivalent to introduce an effective $U(1)$ dynamic gauge-field $b_{\mu}$ into the above action.
\begin{eqnarray}
S'=\int d^{2}xd\tau\sum_{a\sigma}[\bar{\psi}_{a\sigma}(\gamma_{\mu}(\partial_{\mu}-ieA_{\mu}-ie\sigma b_{\mu})
+m_{a\sigma})\psi_{a\sigma}].\nonumber
\end{eqnarray}
It is noted that different spin flavor of electrons have the opposite gauge charge of $b_{\mu}$, which just indicates that the gauge-field $b_{\mu}$ describes the spin fluctuations of ordered magnetic background.

After integrating out Dirac fermions, one can obtain
\begin{eqnarray}
S'_{CS}=\sum_{a\sigma}\int d^{2}xd\tau[\frac{-ie^{2}m_{a\sigma}}{8\pi|m_{a\sigma}|}\epsilon^{\mu\nu\lambda}(A_{\mu}+\sigma b_{\mu})\partial_{\nu}(A_{\lambda}+\sigma b_{\mu})].\nonumber
\end{eqnarray}

For normal antiferromagnetic spin-density wave(N-SDW) states, we have $m_{1\uparrow}=m_{2\downarrow}<0$ and $m_{1\downarrow}=m_{2\uparrow}>0$, thus the resulting action vanishes. In contrast, when we consider the topological antiferromagnetic spin-density wave(T-SDW) states, the effective action reads
\begin{eqnarray}
S'_{CS}=\int d^{2}xd\tau[\frac{-2ie^{2}}{4\pi}\epsilon^{\mu\nu\lambda}(A_{\mu}\partial_{\nu}A_{\lambda}+b_{\mu}\partial_{\nu}b_{\lambda})].
\end{eqnarray}

As argued in Ref. \onlinecite{He2011}, the Chern-Simon term of $b_{\mu}$ gives rise to two gapless chiral modes, which only carry spin degrees of freedom of physical electrons. However, since these two modes do not carry charge degrees of freedom, they are not protected by the conservation of particle number. Therefore, we suspect two chiral modes from the spin fluctuation ($b_{\mu}$) will be gapped and no noticeable spin current could exist in T-SDW states.

\subsection{Spin fluctuation effect at the topological quantum phase transition point and relation to $3D$ $XY$ universal class}
In this subsection, we devote to discuss the Spin fluctuation effect at the topological quantum phase transition point between N-SDW and T-SDW states($m=-3\sqrt{3}t'=-J_{\parallel}/4$) in Sec. \ref{sec2}. Here, the effective mass of Dirac action is $m_{1\uparrow}=m_{2\downarrow}=-J_{\parallel}/2<0$ and $m_{1\downarrow}=m_{2\uparrow}=0$. Thus, $\psi_{1\uparrow},\psi_{2\downarrow}$ are massive and can be safely integrated out while one should not integrate out $\psi_{1\downarrow},\psi_{2\uparrow}$ due to gaplessness of them. Then, following the treatment of last subsection, one obtains
\begin{eqnarray}
&&S''=S_{cs}+S_{1\downarrow}+S_{2\uparrow},\nonumber\\
&&S_{cs}=\int d^{2}xd\tau[\frac{-ie^{2}}{4\pi}\epsilon^{\mu\nu\lambda}(A_{\mu}\partial_{\nu}A_{\lambda}+b_{\mu}\partial_{\nu}b_{\lambda})].\nonumber\\
&&S_{1\downarrow}=\int d^{2}xd\tau[\bar{\psi}_{1\downarrow}\gamma_{\mu}(\partial_{\mu}-ieA_{\mu}+ie b_{\mu})\psi_{1\downarrow}],\nonumber\\
&&S_{2\uparrow}=\int d^{2}xd\tau[\bar{\psi}_{2\uparrow}\gamma_{\mu}(\partial_{\mu}-ieA_{\mu}-ie b_{\mu})\psi_{2\uparrow}],
\end{eqnarray}
where we have integrated out two massive modes, which leads to a Chern-Simon term for $A_{\mu}$ and $b_{\mu}$, respectively. It is interesting to note that the above action resembles to the one in Ref. [\onlinecite{Grover2012}], where such action describes the phase transition between the bosonic integer quantum Hall phases \cite{Senthil2012} and
superfluid phase. (Note that in our case, we have two flavor fermions with opposite gauge charge while in Ref. [\onlinecite{Grover2012}] there exists only one flavor.) In Ref. [\onlinecite{Grover2012}], those bosonic integer quantum Hall phases have quantized Hall conductance $\sigma_{H}=2n\frac{e^{2}}{h}$ with $n=0,1,2,3,...$. (The one with $n=0$ just corresponds to the trivial Bose-Mott insulator.) Obviously, one expects there will be a usual $3D$-$XY$ transition between the bosonic integer quantum Hall phases and the superfluid phase since we have a natural local order parameter if approaching the critical point from the well-defined superfluid phase. Interestingly, it is known that the above fermionic action in fact provides an alternative description for the usual $3D$-$XY$ transition,\cite{Grover2012,Wu1993} thus the correctness of the fermionic effective action is justified and it does not violate our basic physical intuition.

For our case, the T-SDW state has quantized Hall conductance $\sigma_{H}=2\frac{e^{2}}{h}$ while the N-SDW state has vanished quantized Hall conductance. In some sense, one may identify N-SDW and T-SDW states as the superfluid phase and the bosonic integer quantum Hall phases in Ref. [\onlinecite{Grover2012}], respectively, according to distinct features on the quantized Hall conductance. Therefore, we may conclude that the critical behaviors of topological quantum phase transition between N-SDW and T-SDW states could fall into the $3D$-$XY$ universal class though we cannot find a unambiguous bosonic local order parameter for our case. In our view, the similarity between our case and the bosonic integer quantum Hall transition is indeed an interesting new finding and it is desirable to see more examples where some fermionic theories can be dual to certain kinds of bosonic ones.\cite{Poilblanc2012,Poilblanc2008,Kivelson}

\subsection{Kane-Mele-Kondo lattice model}
A careful reader may wonder whether there exists a similar T-SDW state in the Kane-Mele-Kondo lattice model studied in Ref. [\onlinecite{Feng}]. We have studied this model by using the same treatment in the present paper. It is found that the edge states resemble the case in usual 2+1D topological insulator, where the gapless helical edge state exists and the spin Hall conductance is quantized due to the gapless edge state. However, in our case the antiferromagnetic order breaks the time-reversal symmetry, thus the gapless helical edge state will be gapped since it cannot be stable due to impurities and weak interaction without the protection from the time-reversal symmetry.\cite{Hasan2010,Qi2011} Physically, the quantized spin Hall conductance contributed by such gapless edge state will also be destroyed generically and we conclude that no such T-SDW-like states could appear in the Kane-Mele-Kondo lattice model.

\subsection{Bernevig-Hughes-Zhang model with Hubbard-$U$ term}
Recently, it is noted that in Ref. [\onlinecite{Yoshida}], Bernevig-Hughes-Zhang (BHZ) model with an extra Hubbard interaction has been studied by applying the dynamical mean field theory. Those authors find that the Hartree-Fock mean-field theory cannot capture a topological antiferromagnetic phase, which is verified to exist by their dynamical mean-field theory calculation. In contrast to the case of the extended Bernevig-Hughes-Zhang model, in the present paper the T-SDW state is well-captured by our mean-field decoupling and is further inspected by effective Chern-Simon action and chiral edge-state in Sec. \ref{sec2}. We suspect that such difference may result from
the different models one used. It seems that the mean-field decoupling works more effectively in the Kondo lattice-like model than in extended Bernevig-Hughes-Zhang model. Maybe, the usual mean-field approach underestimates the effect of spin-orbit coupling in BHZ model comparing to dynamical mean-field theory calculation and it will be interesting to clarify this point in future work.

\section{conclusion} \label{sec5}

In summary, we have obtained the global ground-state phase diagram of the topological Kondo lattice model on the honeycomb lattice at half-filling by using an extended mean-field decoupling. It is found that besides the well-defined Kondo insulator and the normal antiferromagnetic SDW state, there can exist a novel topological SDW state with quantum anomalous Hall effect. The topological SDW state cannot be distinguished with the normal antiferromagnetic SDW state in terms of the Landau-Ginzburg paradigm, specifically, the symmetry-breaking based classification of states. However, such a novel SDW state can be indeed fully encoded by its low energy effective Chern-Simon action with gapless chiral edge-state.

Moreover, the phase transition between the two kinds of SDW states is a topological quantum phase transition while a first-order quantum phase transition is found between the Kondo insulating state and the antiferromagnetic SDW states. Interestingly, the mentioned topological quantum phase transition can be described by Dirac fermions coupled Chern-Simon gauge-field, which indicates such topological quantum phase transition may fall into the $3D$-$XY$ universal class. It is expected both the global ground-state phase diagram and the novel topological SDW state could be realized by experiments of ultra-cold atoms on the honeycomb optical lattice.\cite{Bloch,Goldman}[We note that authors in Ref. [\onlinecite{Shao}] propose that Haldane model may be realized by using ultracold atoms trapped in an optical lattice. In the topological Kondo lattice we study, the conduction electrons are described by the Haldane model while local electrons couple to conduction electrons via magnetic coupling. Building on the Haldane model, it is expectable to realize the topological Kondo lattice model by adding a magnetic coupling between local electrons and conduction electrons.] We hope the present work may be helpful for further studies on the interplay between conduction electrons and the densely localized spins for the honeycomb lattice.

\begin{acknowledgments}
The authors would like to thank Su-Peng Kou for illuminating and helpful communication on related issues. The work was supported partly by NSFC, the Program for NCET, the Fundamental Research Funds for the Central Universities and the national program for basic research of China.
\end{acknowledgments}

\appendix
\section{Derivation of Chern-Simon action}
Here, we would like to give a brief derivation of the effective Chern-Simon action Eq. (\ref{eq6}) from the Dirac action Eq. (\ref{eq5}).
First, the Dirac action is written as
\begin{eqnarray}
S=\int d^{2}xd\tau\sum_{a\sigma}[\bar{\psi}_{a\sigma}(\gamma_{\mu}(\partial_{\mu}-ieA_{\mu})+m)\psi_{a\sigma}].
\end{eqnarray}
Then, integrating out Dirac fermions one obtains
\begin{eqnarray}
&& S_{eff}=N\ln Det[\gamma_{\mu}(\partial_{\mu}-ieA_{\mu})+m]\nonumber\\
&& =N Tr\ln[\gamma_{\mu}(\partial_{\mu}-ieA_{\mu})+m]\nonumber\\
&& =N Tr[\ln[\gamma_{\mu}\partial_{\mu}+m]+\ln[1-ie(\gamma_{\mu}\partial_{\mu}+m)^{-1}\gamma_{\mu}A_{\mu}]]\nonumber\\
&& \simeq N \int\frac{d^{3}q}{(2\pi)^{3}}A_{\mu}\Pi_{\mu\nu}(q)A_{\nu}
\end{eqnarray}
where $\Pi_{\mu\nu}(q)=\frac{-e^{2}}{2}\int\frac{d^{3}k}{(2\pi)^{3}}Tr[\frac{ik_{\mu}\gamma_{\mu}-m}{m^{2}+k^{2}}\gamma_{\nu}\frac{i(k_{\mu}+q_{\mu})\gamma_{\mu}-m}{m^{2}+(k+q)^{2}}\gamma_{\mu}]
=\frac{-e^{2}}{2}\int\frac{d^{3}k}{(2\pi)^{3}}[\frac{1}{m^{2}+k^{2}}\frac{1}{m^{2}+(k+q)^{2}}][-imq_{\lambda}Tr(\gamma_{\mu}\gamma_{\nu}\gamma_{\lambda})]+...
\simeq\frac{-e^{2}m}{8\pi|m|}\epsilon^{\mu\nu\lambda}q_{\lambda}$ is calculated at one-loop level. We have also used the identity $Tr(\gamma_{\mu}\gamma_{\nu}\gamma_{\lambda})=2i\epsilon^{\mu\nu\lambda}$ with $\gamma_{0}=\tau_{z}$, $\gamma_{1}=\tau_{x}$ and $\gamma_{2}=\tau_{y}$ while $\int\frac{d^{3}k}{(2\pi)^{3}}[\frac{1}{m^{2}+k^{2}}\frac{1}{m^{2}+(k+q)^{2}}]=\frac{\arcsin\left(\frac{|q|}{\sqrt{q^{2}+4m^{2}}}\right)}{4\pi|q|}\simeq\frac{1}{8\pi|m|}$ for $|q|\ll|m|$. Therefore, the effective Chern-Simon action Eq. (\ref{eq6}) is obtained as
\begin{eqnarray}
S_{eff}&&=\int\frac{d^{3}q}{(2\pi)^{3}} N A_{\mu}\frac{-e^{2}m}{8\pi|m|}\epsilon^{\mu\nu\lambda}q_{\lambda}A_{\nu}\nonumber\\
&&=\int d^{2}xd\tau[N e^{2}\frac{-i m}{8\pi|m|}\epsilon^{\mu\nu\lambda}A_{\mu}\partial_{\nu}A_{\lambda}].
\end{eqnarray}

\section{Stability of the gapless chiral Edge-state }
In Sec. \ref{sec2}, the decoupled gapless chiral edge states are described as follows
\begin{eqnarray}
S_{edge}=\sum_{I=1,2}\int dxd\tau\frac{1}{4\pi}[-i\partial_{\tau}\phi_{I}\partial_{x}\phi_{I}+c_{I}\partial_{x}\phi_{I}\partial_{x}\phi_{I}].\nonumber
\end{eqnarray}
To discuss its stability to weak impurities or interaction effect, it is helpful to write down the corresponding fermonic formalism\cite{Giamarchi}
\begin{eqnarray}
H_{edge}^{0}=\sum_{I=1,2}\int dx[\psi^{\dag}_{I}(-ic_{I}\partial_{x})\psi_{I}].
\end{eqnarray}

If the total particle number is conserved, in general, the interacting term will be the following form\cite{Giamarchi}
\begin{eqnarray}
&&H_{int}=H_{1}+H_{2}+H_{3}\nonumber\\
&&H_{1}=g_{1}\int dx[\psi^{\dag}_{1}\psi_{1}\psi^{\dag}_{1}\psi_{1}+\psi^{\dag}_{2}\psi_{2}\psi^{\dag}_{2}\psi_{2}]\nonumber\\
&&H_{2}=g_{2}\int dx[\psi^{\dag}_{1}\psi_{1}\psi^{\dag}_{2}\psi_{2}]\nonumber\\
&&H_{3}=g_{3}\int dx[\psi^{\dag}_{1}\psi^{\dag}_{1}\psi_{2}\psi_{2}+\psi^{\dag}_{2}\psi^{\dag}_{2}\psi_{1}\psi_{1}].
\end{eqnarray}
Obviously, the dangerous $H_{3}$ vanishes due to $\psi_{I}\psi_{I}=\psi^{\dag}_{I}\psi^{\dag}_{I}=0$. $H_{1},H_{2}$ do not vanish but they cannot gap out the edge states since they only correspond to the usual forward scattering.

Besides these interacting terms, the mass term $H_{m}=M\int dx[\psi^{\dag}_{1}\psi_{2}+\psi^{\dag}_{2}\psi_{1}]$, which can result from weak impurity scattering, may be important when considering the stability of the gapless chiral edge-state.\cite{Giamarchi} However, due to the chiral feature of the edge-state,
such mass term does not lead to a gap but a shift of energy (the quasiparticle energy spectrum is $E_{k\pm}=\pm M+ck$), which can be easily compensated by adjusting the chemical potential.
Therefore, we may conclude that the gapless chiral edge-state is at least stable under weak impurities or interaction effect as one expects.


\begin{thebibliography}{99}

\bibitem{Doniach}
S. Doniach, Physica B and C \textbf{91}, 231 (1977).

\bibitem{Sachdev2011}
S. Sachdev, \textit{Quantum Phase Transition}, 2nd ed. (Cambridge University Press, Cambridge, England, 2011).

\bibitem{Rosch}
H. V. L$\ddot{o}$hneysen, A. Rosch, M. Vojta and P. W$\ddot{o}$lfle, Rev. Mod. Phys \textbf{79}, 1015 (2007).

\bibitem{Vojta}
M. Vojta, J Low Temp Phys \textbf{161}, 203 (2010).

\bibitem{Custers1}
J. Custers, P. Gegenwart, H. Wilhelm, K. Neumaier, Y. Tokiwa, O. Trovarelli, C. Geibel, F. Steglich, C.P\'epin and P. Coleman, Nature (London) \textbf{424}, 524 (2003).

\bibitem{Custers2}
J. Custers, P. Gegenwart, C. Geibel, F. Steglich, P. Coleman and S. Paschen, Phys. Rev. Lett. \textbf{104}, 186402 (2010).

\bibitem{Matsumoto}
Y. Matsumoto, S. Nakatsuji, K. Kuga, Y. Karaki, N. Horie, Y. Shimura, T. Sakakibara, A. H. Nevidomskyy and P. Coleman, Science \textbf{331}, 316 (2011).

\bibitem{Senthil2003}
T. Senthil, S. Sachdev and M. Vojta, Phys. Rev. Lett. \textbf{90}, 216403 (2003).

\bibitem{Senthil2004}
T. Senthil, M. Vojta and S. Sachdev, Phys. Rev. B \textbf{69}, 035111 (2004).

\bibitem{Pepin2005}
C. P\'epin, Phys. Rev. Lett. \textbf{94}, 066402 (2005).

\bibitem{Kim2010}
K. S. Kim and C. L. Jia, Phys. Rev. Lett. \textbf{104}, 156403 (2010).

\bibitem{Senthil2010}
T. Grover and T. Senthil, Phys. Rev. B \textbf{81}, 205102 (2010).

\bibitem{Zhong2012e}
Y. Zhong, K. Liu, Y. Q. Wang and H.-G. Luo, Phys. Rev. B \textbf{86}, 115113 (2012).

\bibitem{Tsunetsugu}
H. Tsunetsugu, M. Sigrist and K. Ueda, Rev. Mod. Phys. \textbf{69}, 809
(1997).

\bibitem{Lacroix}
C. Lacroix and M. Cyrot, Phys. Rev. B \textbf{20}, 1969 (1979).

\bibitem{Zhang2000}
G. M. Zhang and L. Yu, Phys. Rev. B \textbf{62}, 76 (2000).

\bibitem{Capponi}
S. Capponi and F. F. Assaad, Phys. Rev. B \textbf{63}, 155114 (2001).

\bibitem{Watanabe}
H. Watanabe and M. Ogata, Phys. Rev. Lett. \textbf{99}, 136401 (2007).

\bibitem{Zhang2010}
G.-B. Li and G.-M. Zhang, Phys. Rev. B \textbf{81}, 094420 (2010).

\bibitem{Zhang2011}
G.-M. Zhang, Y.-H. Su, and Lu Yu, Phys. Rev. B \textbf{83}, 033102 (2011)

\bibitem{Custers2010}
J. Custers, P. Gegenwart, C. Geibel, F. Steglich, P. Coleman and S. Paschen,  Phys. Rev. Lett. \textbf{104}, 186402 (2010).

\bibitem{Wen}
P. A. Lee, N. Nagaosa and X. G. Wen, Rev. Mod. Phys \textbf{78}, 17 (2006).

\bibitem{Saremi}
S. Saremi and P. A. Lee, Phys. Rev. B \textbf{75}, 165110 (2007).

\bibitem{Wen2004}
Xiao-Gang Wen, \textit{Quantum Field Theory of Many-Body Systems} (Oxford Graduate Texts, New York, 2004).

\bibitem{Meng}
Z. Y. Meng, T. C. Lang, S. Wessel, F. F. Assaad, and A. Muramatsu, Nature (London) \textbf{464}, 847 (2010).

\bibitem{Clark}
B. K. Clark, D. A. Abanin and S. L. Sondhi, Phys. Rev. Lett. \textbf{107}, 087204 (2011).

\bibitem{Mezzacapo2012}
F. Mezzacapo and M. Boninsegni, Phys. Rev. B \textbf{85}, 060402(R) (2012).

\bibitem{Zhong2012}
Y. Zhong, K. Liu, Y. Q. Wang and H.-G. Luo, Phys. Rev. B \textbf{86}, 165134 (2012).

\bibitem{Hasan2010}
M. Z. Hasan and C. L. Kane, Rev. Mod. Phys. \textbf{82}, 3045 (2010).

\bibitem{Qi2011}
X.-L. Qi and S.-C. Zhang, Rev. Mod. Phys. \textbf{83}, 1057 (2011).

\bibitem{Rachel}
S. Rachel and K. Le Hur, Phys. Rev. B \textbf{82}, 075106 (2010).

\bibitem{Hohenadler2011}
M. Hohenadler, T.-C. Lang and F. F. Assaad, Phys. Rev. Lett. \textbf{106}, 100403 (2011).

\bibitem{Ruegg2012}
A. R\"{u}egg and G. A. Fiete, Phys. Rev. Lett. \textbf{108}, 046401 (2012).

\bibitem{Mong}
R. S. K. Mong, A. M. Essin and J. E. Moore, Phys. Rev. B \textbf{82}, 075106 (2010).

\bibitem{Essin}
A. M. Essin and V. Gurarie, Phys. Rev. B \textbf{85}, 195116 (2012).

\bibitem{Feng}
X.-Y. Feng, J. Dai, C.-H. Chung and Qimiao Si, arXiv:cond-mat/1206.0979v1 (2012).

\bibitem{Yoshida}
T. Yoshida, R. Peters, S. Fujimoto and N. Kawakami, arXiv:cond-mat/1207.4547v2 (2012).

\bibitem{He2011}
J. He, Y.-H. Zong, S.-P. Kou, Y. Liang and S. Feng, Phys. Rev. B \textbf{84}, 035127 (2011).

\bibitem{He2012}
J. He, Y. Liang and S.-P. Kou, Phys. Rev. B \textbf{85}, 205107 (2012).

\bibitem{Kitaev2008}
A. Kitaev, the Proceedings of the L.D.Landau Memorial Conference ``Advances in Theoretical Physics" (2008)
[See also arXiv:cond-mat/0901.2686v2 (2009)].

\bibitem{Schnyder2008}
A. P. Schnyder, S. Ryu, A. Furusaki and A. W. W. Ludwig, Phys. Rev. B \textbf{78}, 195125 (2008).

\bibitem{Chen2011a}
X. Chen, Z.-G. Gu, Z.-X. Liu and X.-G. Wen, arXiv:cond-mat/1106.4772v4 (2011).

\bibitem{Chen2011b}
X. Chen, Z.-X. Liu and X.-G. Wen, Phys. Rev. B \textbf{84}, 235141 (2011).

\bibitem{Wen2012}
X.-G. Wen, Phys. Rev. B \textbf{85}, 085103 (2012).

\bibitem{Gu2012}
Z.-G. Gu and X.-G. Wen, arXiv:cond-mat/1201.2648v1 (2012).

\bibitem{Levin2012}
M. Levin and Z.-C. Gu, Phys. Rev. B \textbf{86}, 115109 (2012).

\bibitem{Lu2012}
Y.-M. Lu and A. Vishwanath, Phys. Rev. B \textbf{86}, 125119 (2012).

\bibitem{Senthil2012}
T. Senthil and M. Levin, arXiv:cond-mat/1206.1604v1 (2012).

\bibitem{Grover2012}
T. Grover and A. Vishwanath, arXiv:cond-mat/1210.0907v1 (2012).

\bibitem{Lu2012b}
Y.-M. Lu and D.-H. Lee arXiv:cond-mat/1210.0909v1 (2012).

\bibitem{Zhong2012b}
Y. Zhong, K. Liu, Y. F. Wang, Y. Q. Wang and H.-G. Luo, arXiv:cond-mat/1208.1332v2 (2012).

\bibitem{Regnault}
N. Regnault, and B. A. Bernevig, Phys. Rev. X \textbf{1}, 021014 (2011).

\bibitem{Haldane}
F. D. M. Haldane, Phys. Rev. Lett. \textbf{61}, 2015 (1988).

\bibitem{note}
Using the ground-state energy of antiferromagnetic state [Eq. (7)] and Kondo insulator [Eq. (8)] in Ref. [\onlinecite{Zhang2000}], it is easy to check that at the putative QCP $J_{c}=0.58D$, there exists a jump about $0.1$ for the first-order derivative of ground-state energy of antiferromagnetic state and Kondo insulator with respect to the Kondo coupling. Thus, a first-order transition is clearly observed in the mean-field treatment of Ref. [\onlinecite{Zhang2000}].

\bibitem{Continentino}
M. A. Continentino, \textit{Quantum Scaling in Many-Body Systems }(World Scientific Press, Singapore, 2001).

\bibitem{Wu1993}
W. Chen, M. P. A. Fisher, and Y.-S. Wu, Phys. Rev. B \textbf{48},
13749 (1993).

\bibitem{Poilblanc2012}
C. A. Lamas, A. Ralko, D. C. Cabra, D. Poilblanc, and P. Pujol
Phys. Rev. Lett. \textbf{109}, 016403 (2012).

\bibitem{Poilblanc2008}
D. Poilblanc
Phys. Rev. Lett. \textbf{100}, 157206 (2008).

\bibitem{Kivelson}
S. Kivelson
Phys. Rev. B \textbf{39}, 259 (1989).

\bibitem{Bloch}
I. Bloch, J. Dalibard and W. Zwerger, Rev. Mod. Phys. \textbf{80}, 885 (2008).

\bibitem{Goldman}
N. Goldman, A. Kubasiak, A. Bermudez, P. Gaspard, M. Lewenstein, and M. A. Martin-Delgado,
Phys. Rev. Lett. \textbf{103}, 035301 (2009).

\bibitem{Shao}
L.-B. Shao, S.-L. Zhu, L. Sheng, D.-Y. Xing, and Z.-D. Wang,
Phys. Rev. Lett. \textbf{101}, 246810 (2008).


\bibitem{Giamarchi}
T. Giamarchi, \textit{Quantum Physics in One Dimension} (Clarendon,
Oxford, 2005).


\end{thebibliography}
\end{document}